\documentclass[useAMS,usenatbib]{mn2e}
\usepackage{amsmath}
\usepackage{graphicx}
\usepackage{color}

\title[Disc fragmentation vs SMS formation]
{Does disc fragmentation prevent the formation of supermassive stars in protogalaxies?}
\author[K. Inayoshi and Z. Haiman]
{Kohei Inayoshi$^1$
\thanks{E-mail: inayoshi@astro.columbia.edu} 
and Zolt\'an Haiman$^1$
\\
$^1$Department of Astronomy, Columbia University, 550 West 120th Street, New York, NY 10027, USA
}
\newcommand{\msun}{M_\odot}

\newcommand{\zsun}{Z_\odot}
\newcommand{\cc}{{\rm cm}^{-3}}
\newcommand{\mdot}{\dot M}
\newcommand{\msunyr}{M_\odot~{\rm yr}^{-1}}

\begin{document}

\maketitle

\begin{abstract}
Supermassive stars (SMSs; $\ga 10^5~\msun$) formed in the first
protogalaxies with virial temperature $T_{\rm vir}\ga 10^4$ K are
expected to collapse into seeds of supermassive black hole in
the high-redshift Universe ($z\ga 7$).  Fragmentation of the
primordial gas is, however, a possible obstacle to SMS formation.  We
discuss the expected properties of a compact, metal-free, marginally
unstable nuclear protogalactic disc, and the fate of the clumps formed
in the disc by gravitational instability.  Interior to a
characteristic radius $R_{\rm f}={\rm few} \times 10^{-2}$ pc, the disc fragments
into massive clumps with $M_{\rm c}\sim 30~\msun$. The clumps grow via
accretion and migrate inward rapidly on a time-scale of $\sim 10^4$ yr,
which is comparable or shorter than the Kelvin-Helmholtz time
$>10^4$ yr.  Some clumps may evolve to zero-age main-sequence stars and
halt gas accretion by radiative feedback, but most of the clumps can
migrate inward and merge with the central protostar before forming
massive stars.  Moreover, we found that dust-induced fragmentation in
metal-enriched gas does not modify these conclusions unless $Z\ga
3\times 10^{-4}~\zsun$, because clump migration below this metallicity
remains as rapid as in the primordial case.  Our results suggest that
fragmentation of a compact, metal-poor disc can not prevent the
formation of an SMS. 
\end{abstract}


\section{Introduction}

Observations of high-redshift quasars reveal that supermassive
black holes (SMBHs) with mass of $\ga 10^9~\msun$ already exist
as early as redshifts $z\ga 6$
\citep{2006NewAR..50..665F,2007AJ....134.2435W,2011Natur.474..616M}.
Gas accretion and mergers of the remnant black holes (BHs) formed by the collapse of
first-generation stars ($\sim 100~\msun$) have been considered for
producing such SMBHs
\citep[e.g.][]{2001ApJ...552..459H,2003ApJ...582..559V,2007ApJ...665..187L}.
However, various forms of radiative feedback can prevent efficient BH
growth, making it difficult to reach $\ga 10^9~\msun$ within the age
of the high-redshift Universe
\citep{2007MNRAS.374.1557J, 2009ApJ...701L.133A, 2009ApJ...696L.146M,2012ApJ...747....9P,2012MNRAS.425.2974T}.

One possible alternative solution is the rapid formation of
supermassive stars (SMSs; $\ga 10^5~\msun$) and their subsequent
collapse directly to massive BHs in the first galaxies
\citep[e.g.,][]{1994ApJ...432...52L,2006MNRAS.370..289B,2006MNRAS.371.1813L}.
The massive seed BHs, formed by direct collapse, shorten the total
SMBH growth time sufficiently, even in the presence of subsequent
radiative feedback (e.g. \citealt{th2009,2012ApJ...745L..29D}).

SMSs can form out of primordial gas in massive `atomic cooling'
haloes with virial temperatures $T_{\rm vir}\ga 10^4$K, if H$_2$
formation and line cooling are prohibited through the pre-stellar
collapse.  Possible mechanisms to suppress H$_2$ formation are
photo dissociation by far-ultraviolet (FUV) radiation
\citep{2001ApJ...546..635O,2003ApJ...596...34B,2008MNRAS.391.1961D,
2009MNRAS.396..343R,2010MNRAS.402.1249S,
2011MNRAS.416.2748I,2011MNRAS.418..838W,2013MNRAS.428.1857J} and
collisional dissociation
\citep{2012MNRAS.422.2539I,2014MNRAS.439.3798F}.  In the absence of
${\rm H_2}$, the primordial gas remains warm ($\sim8,000$K) and may
collapse monolithically without strong fragmentation
\citep{2003ApJ...596...34B,2010MNRAS.402.1249S,
  2013MNRAS.433.1607L,2014arXiv1404.4630I}.  The resulting central
protostar can grow via accretion at a high rate of $\ga 0.1~\msunyr$.
If the accretion rate maintains such a high value, the embryonic
protostar evolves to a SMS with mass of $\ga 10^5~\msun$.  The SMS
collapses as a whole to a single BH either directly
\citep{2002ApJ...572L..39S}, or via the intermediate stage of a close
binary BH \citep{2013PhRvL.111o1101R}, through a general relativistic
instability \citep[e.g.,][]{1971reas.book.....Z,1983bhwd.book.....S}.
The massive remnant BH is then a promising seed that can grow into one
of the observed SMBHs at $z\approx 6-7$.

In the scenario above, a major unresolved question is whether rapid
accretion will continue unabated through the protostellar evolution.
Recent high-resolution simulations by \cite{2014MNRAS.439.1160R}
suggest that the compact nuclear accretion disc, surrounding a central
embryonic protostar, is gravitationally unstable.  Thus,
self-gravitating clumps are expected to form during the early stages
of the accretion phase.  Such efficient fragmentation could prevent
the rapid growth of the central protostar, and preclude eventual SMS
formation.  This is analogous to the fragmentation of gravitationally
unstable discs in the cores of lower mass `minihaloes', which had
been suggested to reduce the characteristic masses of Population III (Pop III)
stars \citep{2010MNRAS.403...45S, greif+11}.  Moreover, if the gas is slightly polluted by
heavy elements ($Z\la 10^{-4}~\zsun$), dust cooling can decrease the
temperature and induce the fragmentation \citep{2008ApJ...686..801O},
which would be a further obstacle to SMS formation.  Because of this,
the existence of the pristine, metal-free gas has often been
considered as a necessary condition of forming a SMS in
semi-analytical models \citep[e.g.,][]{2012MNRAS.425.2854A,2014MNRAS.442.2036D}.
We note that even very efficient fragmentation may not prevent accretion 
on to a central point source \citep[e.g.,][]{2007ApJ...671.1264E} and
the rapid inward migration of the fragments, on a time-scale comparable 
to the orbital period, may help its growth further 
\citep[e.g.,][]{2010MNRAS.404.2151C}.
In our context, however, even a slow down of the accretion rate could bring it 
below the critical value required for SMS formation.

In this paper, motivated by the above, we discuss the expected properties of a compact,
marginally unstable nuclear protogalactic disc, and the fate of the
clumps formed in the disc by gravitational instability.  Using
analytical models, we argue that despite fragmentation, the growth of
a SMS remains the most likely outcome. The reason for
this conclusion is the rapid inward migration of the fragments, and
their merger with the central protostar.

The rest of this paper is organized as follows.  In \S~\ref{sec:2}, we
describe the basic model of the fragmenting disc and of the clumps
formed in the disc.  In \S~\ref{sec:fate}, we discuss the fate of the
clumps, considering various important processes: migration, accretion,
contraction, and star-formation by reaching the zero-age main sequence
(ZAMS).  We also consider the possibility that radiative feedback from
massive stars, formed from the clumps, may halt gas accretion on to the
whole system.  In \S~\ref{sec:metal}, we discuss whether
metal pollution and associated dust cooling could prohibit SMS
formation.  Finally, we discuss our results and summarize our
conclusions in \S~\ref{sec:conc}.


\section{Fragmentation of the accretion disc around a supermassive star}
\label{sec:2}

\subsection{Basic equations}

We consider the properties of a disc formed after the collapse of
primordial gas inside an atomic cooling halo,
when molecular hydrogen formation is suppressed.
Since the parent gas
cloud is hot ($T\sim 8000$ K), the accretion rate on to the disc is
high:
\begin{equation}
\mdot_{\rm tot} \sim \frac{c_{\rm s}^3}{G}\sim 0.1~\msunyr \left(\frac{T}{8000~{\rm K}}\right)^{3/2},
\end{equation}
where $c_{\rm s}$ is the sound speed and $G$ is the gravitational
constant \citep{1977ApJ...214..488S, 1986ApJ...302..590S}.  The
accretion disc around the central protostar can become unstable
against self-gravity.  To understand the stability of the disc,
Toomre's parameter \citep{1964ApJ...139.1217T} defined by
\begin{equation}
Q=\frac{c_{\rm s}\Omega}{\pi G\Sigma}
\label{eq:Q}
\end{equation}
is often useful, where $\Omega$ is the orbital frequency and $\Sigma$
is the surface density of the disc.  In the marginal case of $1\la Q
\la 2$, strong spiral arms are formed in the disc.  The spiral arms
redistribute angular momentum and heat the disc by forming shocks
\citep[e.g.,][]{2005ApJ...633L.137V,2006ApJ...650..956V}.
The resulting disc is self-regulated to the marginal state. 
Thus, we here assume $Q\simeq 1$.

Since the gas temperature in the atomic cooling halo is kept near
$\sim 8000$ K, the external accretion rate on to the disc, from larger
radii, is also nearly constant.  We estimate the surface density of
the self-regulated disc by assuming that it is in steady state, and
that it has an effective viscosity $\nu$ (arising from gravitational
torques),
\begin{equation}
\Sigma=\frac{\mdot_{\rm tot}}{3\pi \nu}.
\label{eq:continue}
\end{equation}
The scale-height of the disc is estimated from vertical hydrostatic
balance,
\begin{equation}
H=\frac{c_{\rm s}}{\Omega}. 
\label{eq:scaleheight}
\end{equation}
We define the particle number density as $n\equiv\Sigma /(2m_{\rm
  p}H)$, where $m_{\rm p}$ is the proton mass.

\begin{figure*}
\vspace{1mm}
\begin{center}
\begin{tabular}{ccc}
 \begin{minipage}{0.33\hsize}
  \begin{center}
  \vspace{-1mm}
   \includegraphics[width=54mm]{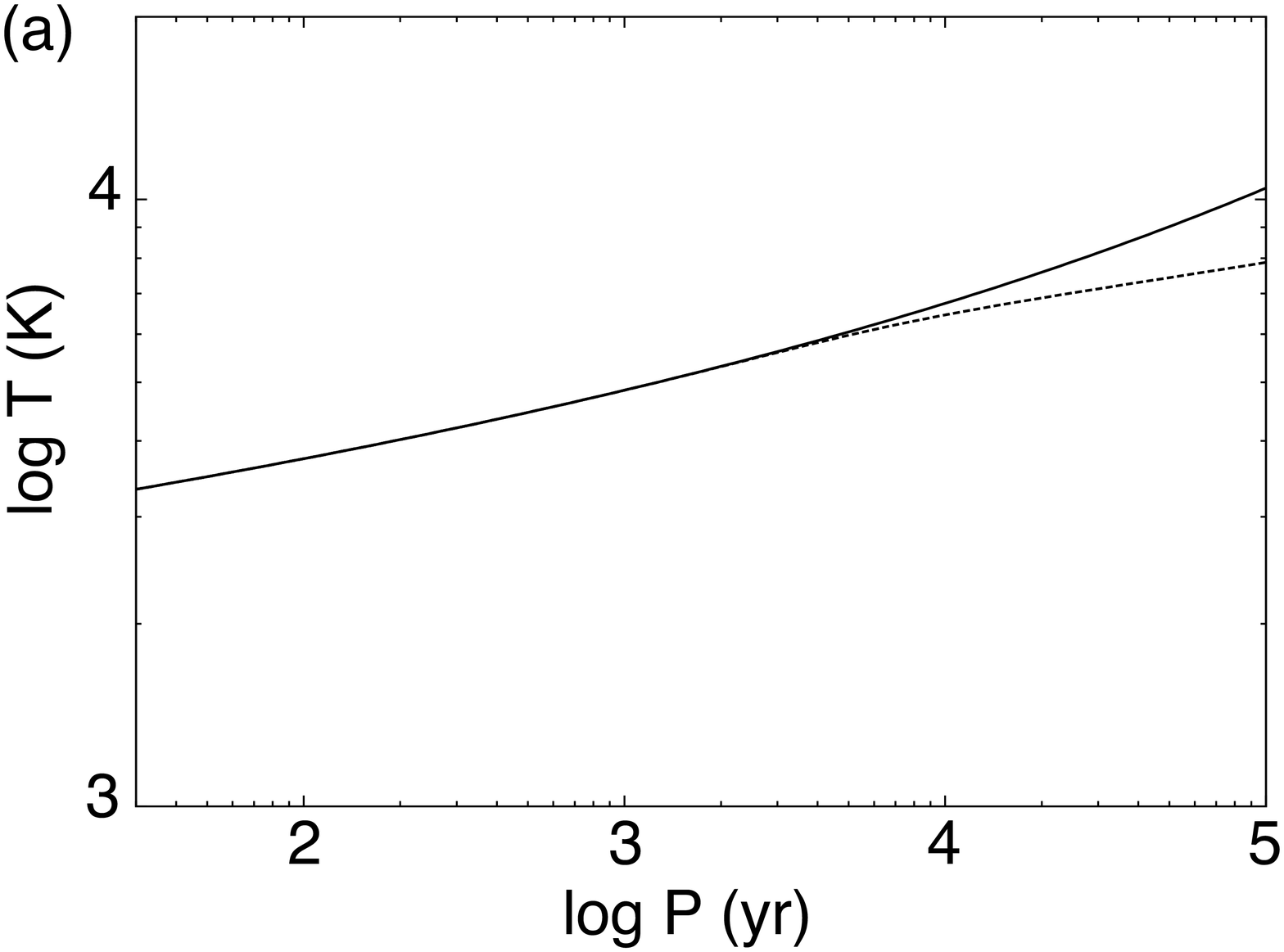}
  \end{center}
 \end{minipage}
\begin{minipage}{0.32\hsize}
 \begin{center}
   \includegraphics[width=54.5mm]{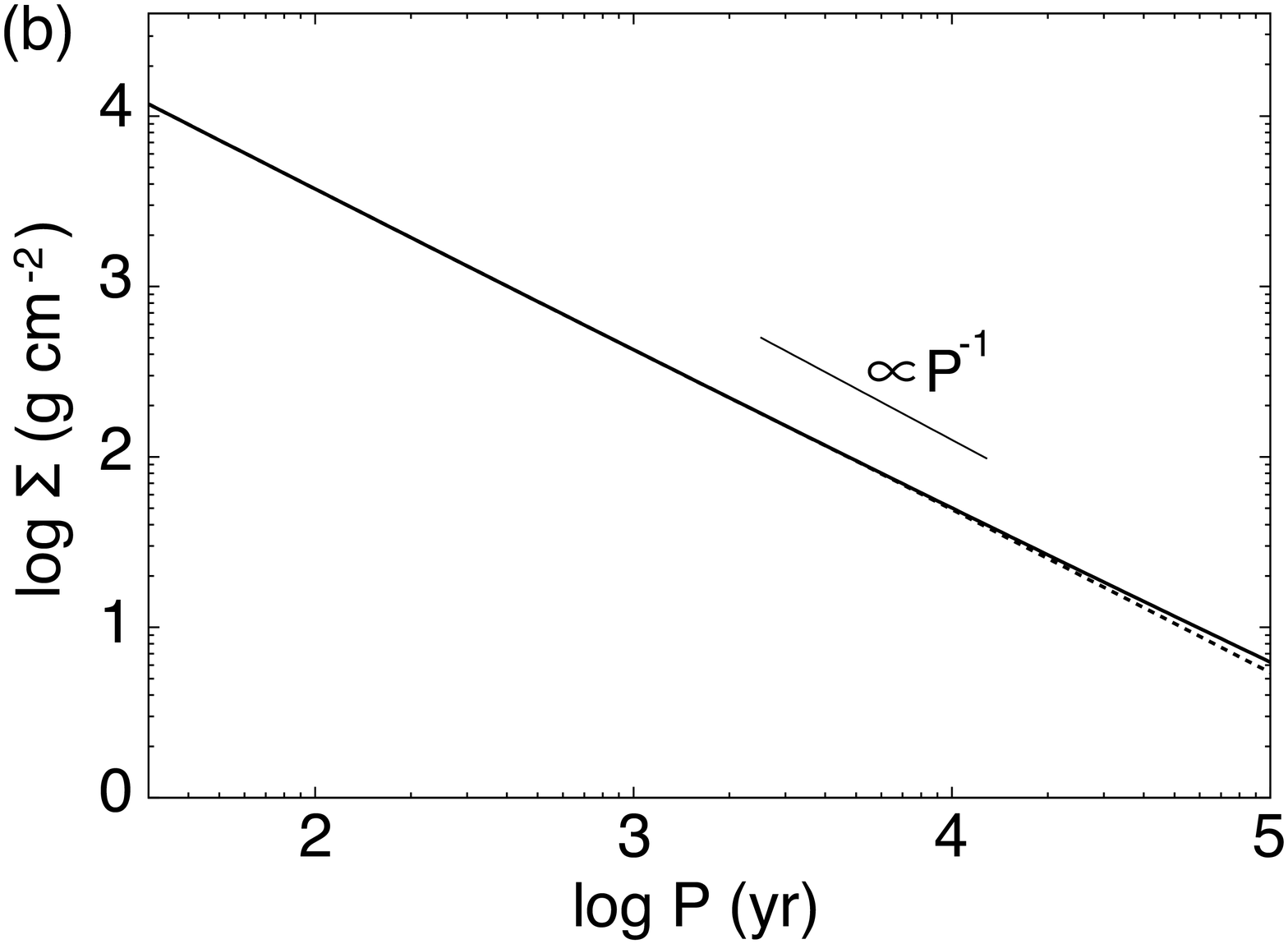}
  \end{center}
 \end{minipage}
\begin{minipage}{0.32\hsize}
 \begin{center}
   \includegraphics[width=55mm]{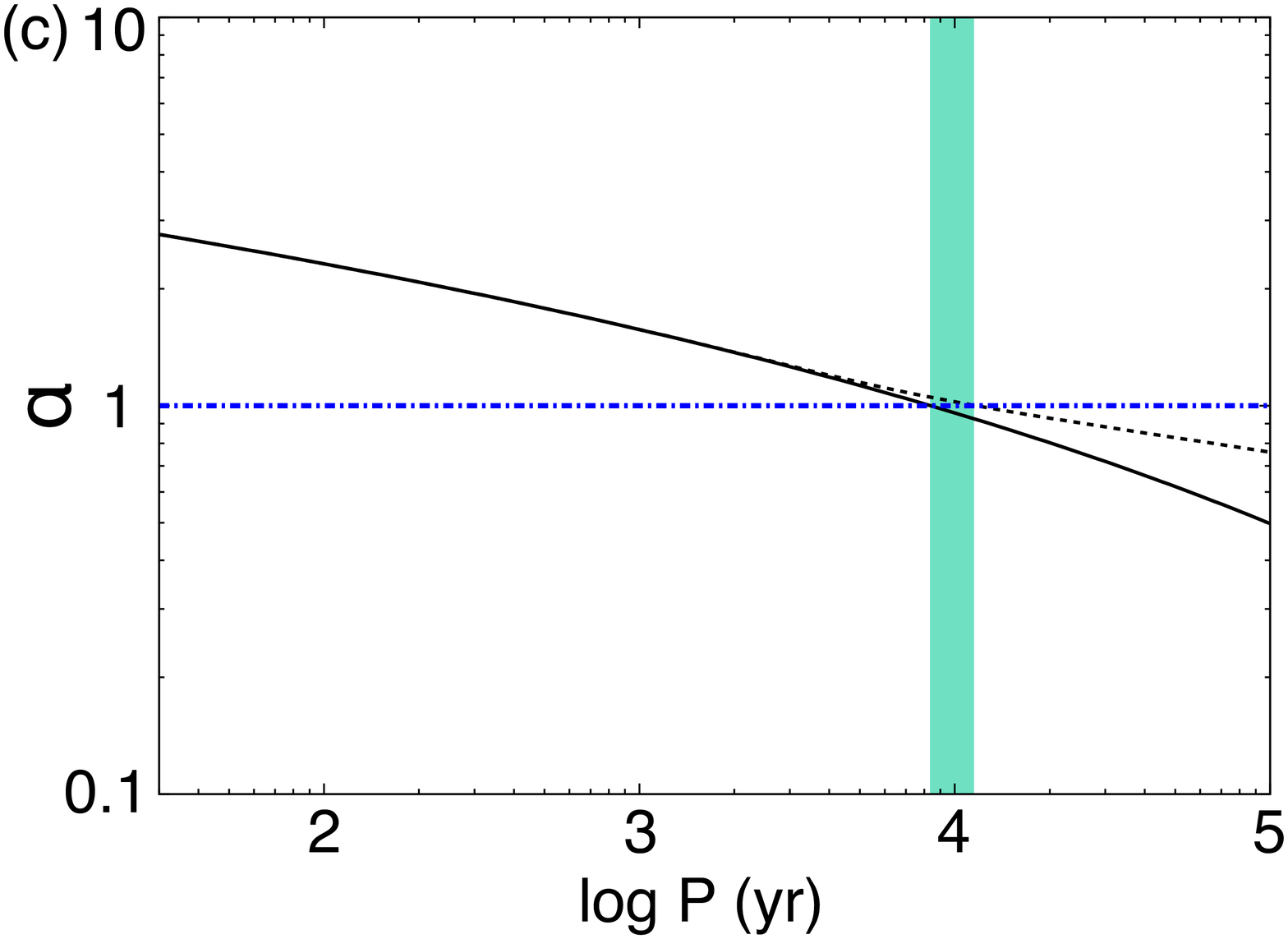}
  \end{center}
 \end{minipage}
\end{tabular}
\caption{Radial profiles of disc quantities, shown as a function of
  the orbital period: (a) temperature, (b) surface density, and (c)
  viscous parameter.  The solid curves show profiles with H$^-$
  free-bound emission as the only radiative cooling process.  The
  dashed curves additionally include Ly$\alpha$ emission.  The
  horizontal dot-dashed line in panel (c) indicates the critical value
  of the viscous parameter for fragmentation, $\alpha_{\rm f}=1$.  The
  shaded bar marks the location of the fragmenting radius at
  $P_{\rm f}=2\pi/\Omega_{\rm f}=10^4$ yr (or $\Omega_{\rm f}=2\times 10^{-11}$
  s$^{-1}$).  }
  \label{fig:profile}
\end{center}
\end{figure*}

To determine the thermal state of the disc we assume that it is in
equilibrium, i.e. we balance heating and radiative cooling
($Q_+=Q_-$).  In particular, we consider viscous heating by turbulence
and spiral shocks,
\begin{align}
Q_+&=\frac{9}{4}\nu \Sigma \Omega ^2,
\end{align}
and radiative cooling
\begin{align}
Q_-&=2H\Lambda,
\end{align}
where $\Lambda$ is the cooling rate in units of erg s$^{-1}$
cm$^{-3}$.  For the cooling process, we consider H$^-$ free-bound
emission, which is the dominant channel in the thermal evolution at
$n\ga 10^8~\cc$ for a warm primordial gas in an atomic cooling halo.
The form of the cooling rate is given in \S~\ref{sec:chem}.

Finally, we adopt the standard $\alpha$-prescription
\citep{1973A&A....24..337S} as a model for the viscosity by
gravitational torques:
\begin{equation}
\nu = \alpha c_{\rm s}H,
\label{eq:viscous}
\end{equation}
where $\alpha$ is the viscous parameter.  From hydrodynamical
simulations of a self-gravitating disc, the disc is found to be
susceptible to fragmentation if the viscous parameter exceeds a
critical threshold value, $\alpha \ga \alpha_{\rm f}$
\citep[e.g.,][]{2001ApJ...553..174G}.  \cite{2005MNRAS.364L..56R} have
studied fragmentation conditions for various specific heat ratios, and
obtained $\alpha_{\rm f}\sim 0.06$.  However, the critical value
depends on both initial conditions and on numerical resolution
\citep{2011MNRAS.410..559M, 2011MNRAS.411L...1M}.  Recently,
\cite{2012ApJ...746..110Z} have considered mass loading from an
infalling envelope, realistic radiative cooling, and radiative
trapping of energy inside clumps, and suggested that the critical
value for fragmentation is around $\alpha_{\rm f}\sim 1$. Moreover,
numerical simulations of fragmentation of discs with relatively high
accretion rates ($\mdot_{\rm tot}\sim 10^{-4}-10^{-2}~\msunyr$), in
the context of present-day massive star formation \citep[e.g.,
][]{2007ApJ...656..959K} and Pop III star formation
\citep{2011Sci...331.1040C} have shown that the effective viscous
parameter arising from gravitational torques is $\sim 0.1-1$.  Based
on these results, we here set $\alpha _{\rm f}=1$ as our fiducial
value, and analyze disc fragmentation using the condition of $\alpha
\geq \alpha_{\rm f}$ following previous analytical works
\citep[e.g.,][]{2005ApJ...621L..69R, 2007MNRAS.374..515L}.

\subsection{Radiative cooling and chemistry}
\label{sec:chem}

In dense ($n\ga 10^8~\cc$) and warm ($3000\la T\la 8000$ K) primordial
gas, the dominant cooling processes is the free-bound emission of
H$^-$ ions (H + e$^-$ $\rightarrow$ H$^-$ + $\gamma$;
\citealt{2001ApJ...546..635O,2014arXiv1404.4630I}). The cooling rate,
in units of erg s$^{-1}$ cm$^{-3}$, is given by
\begin{equation}
\Lambda =\lambda(T)n^2x_{\rm e},
\end{equation}
where $x_{\rm e}(\ll1)$ is the free-electron fraction and
$\lambda(T)\simeq 10^{-30}~ T$ (with $T$ in units of Kelvin).  
In such a warm gas, prior to protostar formation, 
the electron fraction is determined by the balance between
recombination (H$^+$ + e$^-$ $\rightarrow$ H + $\gamma$) and
collisional ionization (2H $\rightarrow$ H$_2^+$ + e$^-$).
The evolution of the electron fraction follows the equation
\begin{equation}
\frac{dx_{\rm e}}{dt}=-\alpha_{\rm rec}nx_{\rm e}^2 + \alpha_{\rm ci}n.
\end{equation}
At the densities of interest here, the electron fraction is in equilibrium,
and given simply by 
\begin{equation}
x_{\rm e}=\sqrt{\frac{\alpha_{\rm ci}}{\alpha_{\rm rec}}}.
\end{equation}
From Appendix \ref{sec:appA}, we obtain the specific value of
\begin{align}
x_{\rm e}&\simeq f(T)~n^{1/2} \exp(-\epsilon/2T)\label{eq:x_e}\\\nonumber
&\simeq 5.0\times 10^{-11}~T^{1.2}~n^{1/2} \exp(-\epsilon/2T).
\end{align}

\subsection{Radial profile of the disc}
\label{sec:disc}

We next obtain the properties of the marginally fragmenting disc,
using the energy conservation equation ($Q_+=Q_-$).  Using
Eq.~(\ref{eq:Q}), the particle density can be written by
\begin{equation}
n=\frac{\Sigma \Omega}{2m_{\rm p}c_{\rm s}}=\frac{\Omega ^2}{2m_{\rm p}\pi G}.
\end{equation}
Then, we can solve the energy conservation equation with respect to $\Omega$:
\begin{equation}
\Omega ^2=
\frac{3\mdot_{\rm tot}}{8\pi}
\frac{(2\pi G m_{\rm p})^{5/2}}{c_{\rm s} \lambda(T)f(T)}
\exp(\epsilon/2T).
\end{equation}

We present the profiles of the temperature, surface density, and
viscous parameter as a function of $\Omega$ for the case of
$\mdot_{\rm tot} =0.1~\msunyr$ in Figure~\ref{fig:profile}.  As
described above, we assume that fragmentation occurs efficiently at
the radii where $\alpha \ga \alpha_{\rm f}\simeq 1$.  From
Figure~\ref{fig:profile}(c), we find that fragmentations occur in the
central regions where the orbital period is shorter than $10^4$ yr.
Within the fragmenting region, the surface density is approximately
given by $\Sigma =\Sigma_{\rm f}(\Omega/\Omega_{\rm f})$, where
$\Sigma_{\rm f}=50$ g cm$^{-2}$ and $\Omega_{\rm f}=2\times 10^{-11}$
s$^{-1}$, respectively.  
During the earliest stages, i.e. prior to the formation of any central
protostar embryo, or immediately following it, the disc mass dominates
the central protostellar mass.  We can then obtain the radius within
which the disc fragments effectively,
\begin{equation}
R_{\rm f}=\frac{2\pi G\Sigma_{\rm f}}{\Omega_{\rm f}^2}\simeq 2\times 10^{-2}~{\rm pc}.
\label{eq:r_f_early}
\end{equation}
Moreover, during this stage, the profiles of the surface density, disc
mass, and number density can be written as functions of the radial
distance $R$ from the central protostar:
\begin{equation}
\Sigma = \Sigma_{\rm f}\left(\frac{R}{R_{\rm f}}\right)^{-1},
\label{eq:sig0}
\end{equation}
\begin{equation}
M_d \simeq 430~\msun \left(\frac{R}{R_{\rm f}}\right),
\label{eq:md0}
\end{equation}
and
\begin{equation}
n \simeq 6\times 10^8~\cc \left(\frac{R}{R_{\rm f}}\right)^{-2},
\label{eq:numdens}
\end{equation}
respectively.
The typical mass of the clumps formed at $\simeq R_{\rm f}$
is estimated as
\begin{equation}
M_{\rm c}\simeq \Sigma_{\rm f} H_{\rm f}^2\simeq 30~\msun.
\label{eq:M_c}
\end{equation}
These disc profiles and the clump mass are in good agreement with the
highest resolution results in the numerical simulations by
\cite{2014MNRAS.439.1160R}, lending credence to our simplified `toy'
model.

In the above, we have assumed that the mass of the central protostar
is negligible.  However, the central protostar grows via rapid
accretion and having a central point source can then modify the disc
structure.  After $\ga 10^4$ yr, the gas within $R_{\rm f}$ accretes
on to the central protostar, and the protostellar mass exceeds the disc
mass within $R_{\rm f}$.  The fragmentation radius (defined by
$\alpha(R_{\rm f})=1$)
remains roughly constant for the first $\sim {\rm a~few}~\times 10^4$ yr, after which it
begins to move outward slowly, according to
\begin{align}
R_{\rm f}&=\left(\frac{GM_\ast}{\Omega_{\rm f}^2}\right)^{1/3}\label{eq:rf_late}
\\\nonumber
&\simeq 5\times 10^{-2}~{\rm pc}\left(\frac{M_\ast}{10^4~\msun}\right)^{1/3}
\propto t^{1/3},
\end{align}
where $M_\ast$ is the mass of the protostellar embryo at the center,
and we have assumed that the central accretion rate remains constant
over time.
In the regions where $M_\ast>M_{\rm d}$, the orbital frequency is
proportional to $R^{-3/2}$ and thus the surface density and mass of
the disc are given by
\begin{equation}
\Sigma =\frac{\Sigma_{\rm f}}{\Omega_{\rm f}}\sqrt{\frac{GM_\ast}{R^3}}
\label{eq:sig1}
\end{equation}
and
\begin{equation}
M_d =4\pi \frac{\Sigma_{\rm f}}{\Omega_{\rm f}}\sqrt{GM_\ast R},
\label{eq:md1}
\end{equation}
respectively. The radius where $M_\ast=M_{\rm d}$ is given by
\begin{equation}
R_{\rm g\ast} =\frac{M_\ast \Omega_{\rm f}^2}{16\pi^2 G\Sigma_{\rm f}^2}\propto t.
\end{equation}
The condition $R_{\rm f}\la R_{\rm g\ast}$ is satisfied when the
protostellar mass exceeds $\simeq 3\times 10^3~\msun$, which corresponds to 
$t\simeq 3\times 10^4~{\rm  yr}~(M_\ast/3\times 10^3~\msun )(\mdot_{\rm tot}/0.1~\msunyr )^{-1}$.
For convenience in the following section (\S~\ref{sec:fate}), we define the transition epoch as $t_{\rm g\ast}=10^5$ yr
neglecting the difference of the factor of 3 because the ratio of $R_{\rm g\ast}/R_{\rm f}\propto t^{2/3}$ 
does not change significantly in the range $3\times 10^4<t<10^5$ yr. 
Therefore, we use Eq.~(\ref{eq:sig0}) and (\ref{eq:md0}) for $t<t_{\rm g\ast}$
and Eqs.~(\ref{eq:sig1}) and (\ref{eq:md1}) for $t\geq t_{\rm g\ast}$
(hereafter the `late phase') for the radial profile of the disc,
respectively.


\section{Fate of the central protostar and clumps}
\label{sec:fate}

In the fragmenting disc, the evolution of the clumps is determined by
several physical processes, including migration, accretion, and
evolution to ZAMS stars.  In this section, we consider these
processes, and discuss the fate of the clumps formed at the
fragmentation radius.

\subsection{Clump migration}
\label{sec:mig}

The clumps formed at $R_{\rm f}$ interact with the disc through
gravitational torques, which results in their orbital evolution, i.e.
`migration'.  We first discuss the time-scales for limiting cases of
inward migration applicable to planets embedded in smooth discs; this
serves as a useful reference.  We then discuss the complications and
uncertainties for protogalactic discs, arising from the fact that the
disc is clumpy, and the nominal migration rate is very rapid.

Limiting cases of idealized migration are well-known in the theory of
disc-planet interactions.  One is the case when the planet mass is so
small that the perturbation induced in the disc is in the linear
regime.  The density waves excited in the disc by the planet exert
torques on the planet through resonances, and thus promote the
migration (so-called Type~I migration;
e.g. \citealt{1980ApJ...241..425G,1986Icar...67..164W}).  The
characteristic time-scale of Type~I migration is
\begin{equation}
t_{\rm mig,I}=\frac{1}{4Cq\mu}\left(\frac{H}{R}\right)^2\frac{2\pi}{\Omega},
\label{eq:typeI}
\end{equation}
where $C=3.2+1.468~\xi$, $\xi$ is the power of the surface density
profile $\Sigma(R)$, $q=M_{\rm c}/M_\ast$, and $\mu=\pi \Sigma
R^2/M_\ast$ \citep{2002ApJ...565.1257T}.  A second case is when the
planet mass is massive enough to clear a gap in the disc (but still
less massive than the local disc).  Then, the planet migrates inward
on approximately the discs's viscous time-scale (so-called Type~II
migration; e.g. \citealt{1986ApJ...309..846L,1997Icar..126..261W}):
\begin{equation}
t_{\rm mig,II}\simeq t_{\rm vis}
=\frac{1}{3\pi \alpha}\left(\frac{R}{H}\right)^2 \frac{2\pi}{\Omega}.
\label{eq:typeII}
\end{equation}

The theoretical basis for these time-scales is derived from idealized
setups, assuming point-like planets and a smooth disc profile.
Whether these results are applicable for an extended clump, in a
gravitationally unstable and lumpy disc, as we discuss in this paper
is poorly known.  Nevertheless, recent numerical simulations have
provided some useful information.  In particular, low-mass planets
embedded in a turbulent disc have been found to migrate inwards,
despite the stochastic kicks due to the turbulent density
fluctuations, essentially on the time-scale of Type I migration
\citep{2011MNRAS.416.1971B, 2011ApJ...737L..42M}.  If the clump grows
via accretion and opens a gap during its migration in the
gravitationally unstable disc, a transition between Type~I and Type~II
migration occurs and the migration speed slows down
\citep{2012ApJ...746..110Z}.

From the disc properties discussed in \S~\ref{sec:disc}, we expect
that the characteristic migration time shifts from that of Type~II to
that of Type~I, because the mass ratio $q$ decreases as the central
protostar (and the mass of the corresponding steady-state disc) grows.
We therefore estimate the transition epoch and the migration time for
the late phase ($M_\ast \ga 10^4~\msun$) using
Eq.~(\ref{eq:typeI}).  The gap-opening condition has been
investigated by many authors
\citep[e.g.,][]{1979MNRAS.186..799L,2002ApJ...572..566R,2006Icar..181..587C,duffellmacfadyen2013},
but again, remains poorly known for gravitationally unstable and
clumpy discs.  To give a conservative estimate, we here do not discuss
the gap-opening condition, but compare the time-scales of Type~I and II
migration.  The ratio of the two time-scales at $R_{\rm f}$ (Eq.~\ref{eq:rf_late}) 
is independent of the central mass, and is given by
\begin{equation}
\frac{t_{\rm mig,I}}{t_{\rm mig,II}}\simeq \frac{0.4}{q}\left(\frac{H}{R}\right)^{3}
\simeq 2.0\left(\frac{M_{\rm c}}{30~\msun}\right)^{-1},
\end{equation}
where we have used the condition $Q=1$.  Since the clumps grow via
accretion at a typical rate of at least $\mdot_{\rm c}\sim {\rm a~few}
\times 10^{-3}~\msunyr$ (see \S~\ref{sec:kh}), the clump
migrates towards the central protostar on a time-scale comparable or
shorter than that in the Type~II case, independent of the presence or
absence of a gap.  In the following, we therefore adopt $t_{\rm mig,
  II}$ as a conservative choice for the migration time-scale.

We first estimate the migration time for the case that the
self-gravity of the disc within $R_{\rm f}$ dominates the protostellar
gravity (i.e. $t<t_{\rm g\ast}$).  At this stage, the mass of
the central protostar is $\la 10^4~\msun$.  From
Eqs.~(\ref{eq:md0}) and (\ref{eq:M_c}), the clump mass is
initially smaller than the local disc mass ($M_{\rm c}\simeq
0.1~M_d$).  In the disc-dominated case, the orbital decay time of the
clump ($t_{\rm mig}=-R/\dot R$) is approximately given by the viscous
time (Eq.~\ref{eq:typeII}; see also \citealt{duffell+2014} who
find a factor of few faster migration in the case they studied).
Since $H/R=Q/2\simeq 0.5$ at $R_{\rm f}$, $t_{\rm mig}\sim 0.4
(2\pi/\Omega_{\rm f})\sim 4\times 10^3$ yr.  Thus, we find that the
clump orbit decays slightly faster than the orbital period. This
follows essentially from our assumption of $\alpha_{\rm f}=1$,
and leads to a complication: Eq.~(\ref{eq:typeII}) for the
Type~II migration rate is strictly valid only for $t_{\rm mig}>
2\pi/\Omega$ because the torques exerted on the clump are assumed to
be averaged over an orbit (at a fixed radius).  Despite this
complication, we can still safely conclude $t_{\rm mig}\la
2\pi/\Omega$. This is because assuming a slower migration ($t_{\rm
  mig}\ga 2\pi/\Omega$) would justify the use of orbit-averaged
Type~II torques, which would then yield the contradiction $t_{\rm
  mig}< 2\pi/\Omega$.

\begin{figure}
\begin{center}
\includegraphics[height=58mm,width=82mm]{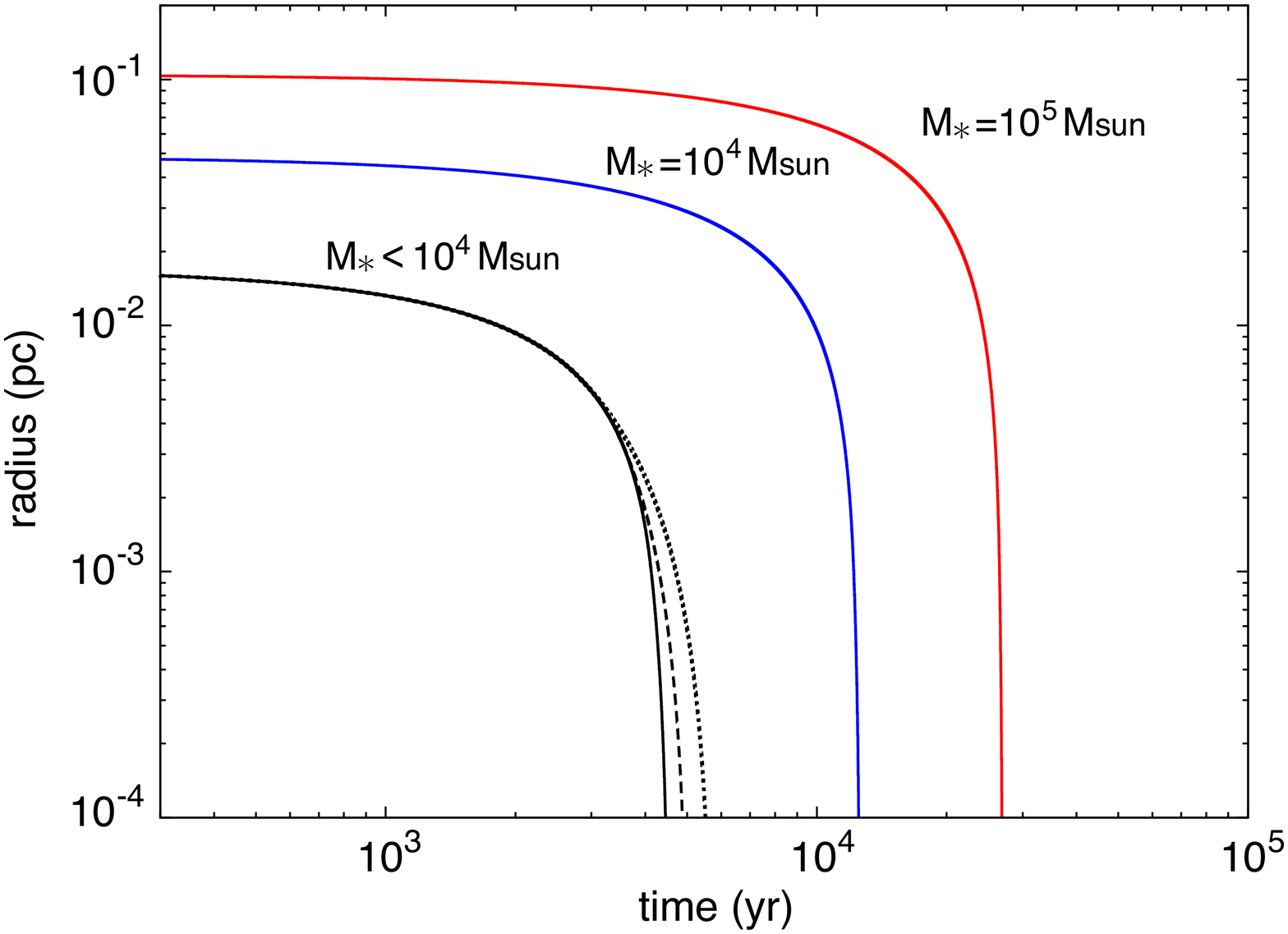}
\end{center}
\caption{The decay time of the clump's orbit, for the case of $M_\ast
  <10^4~\msun$ ($M_\ast<M_{\rm d}$; $t<t_{\rm g\ast}$), $M_\ast =10^4$ and $10^5~\msun$
  ($M_\ast>M_{\rm d}$; $t>t_{\rm g\ast}$), respectively.  In each case, the initial
  position of the clump is set to the fragmentation radius $R_{\rm
    f}$.  In the first ($M_\ast<10^4~\msun$) case, we show the
  slow-down of the migration expected if the clump grows by accretion
  at a rate of $f\mdot_{\rm tot}$ with $f=0$ (solid), $0.2$ (dashed),
  and $0.5$ (dotted).}
\label{fig:mig}
\end{figure}

When the clump migrates to $\simeq 0.1~R_{\rm f}$, its mass becomes
comparable to the local disc mass (neglecting for now any growth of
either the disc or the clump during the migration).
Within $\sim 0.1~R_{\rm f}$, the migration speed slows down because
the disc outside the clump can no longer absorb the orbital angular
momentum of the clump \citep{1995MNRAS.277..758S}.
In the clump-dominated case, the migration time is somewhat modified
as
\begin{equation}
t_{\rm mig}\simeq q_{\rm B}^{-k}t_{\rm vis},
\end{equation}
where
\begin{align}
q_{\rm B}&=\frac{(1+q)\mdot_{\rm tot}}{M_{\rm c}}t_{\rm vis}\\\nonumber
&\simeq 1.1~\left(\frac{1+q}{1.1}\right)\left(\frac{M_{\rm c}}{30~\msun}\right)^{-1}
\left(\frac{\mdot_{\rm tot}}{0.1~\msunyr}\right)\left(\frac{t_{\rm vis}}{300~{\rm yr}}\right)
\end{align}
and
\begin{align}
k=1-\left(1+\frac{\partial \ln \Sigma}{\partial \ln \mdot_{\rm tot}}\right)^{-1}\simeq 0.4.
\end{align}

In Figure~\ref{fig:mig}, we show the evolution of the orbital radius
of the clump (black lines: $M_\ast<10^4~\msun$).  The horizontal axis
is the time from the onset of the orbital decay at $R_{\rm f}$.  To see the
effect of the slow down, we assume that the clump grows at the
accretion rate of $f\mdot_{\rm tot}$ with $f=0$ (solid curve), $0.2$
(dashed), and $0.5$ (dotted), respectively.  In each case, the clump's
orbit decays within $10^4$ yr, even if we consider the slow down of
the migration.

Next, we consider the case that the protostellar gravity exceeds the
self-gravity of the disc within $R_{\rm f}$ (i.e. $t\geq t_{\rm g\ast}$).  
In Figure~\ref{fig:mig}, we show the corresponding
orbital evolution for $M_\ast =10^4$ (blue middle curve) and
$10^5~\msun$ (red right most curve) as examples of this stage.
In both the cases, we do not consider the clump growth because
the initial clump mass is much smaller than the disc mass within 
$R_{\rm f}$ and thus the slow-down effect works after the orbit
has decayed by more than two orders of magnitude.
For $M_\ast =10^5~\msun$, the fragmentation radius moves out to $\sim 0.1$ pc. 
In this case, the decay time becomes longer than the orbital
period, and the clump could evolve into a normal massive star, rather
than an SMS. However, this scenario is realized only by assuming that
a SMS with $M_\ast \sim 10^5~\msun$ has already grown at the center of
the disc.

\subsection{Clump accretion and evolution}
\label{sec:kh}

In the fragmenting disc, many clumps are formed within $R_{\rm f}$ and
their orbits decay through interaction with the accretion disc.  Since
the typical density of the clumps is $\sim 10^9~\cc$ (Eq.~\ref{eq:numdens})
and since they are optically thin to the H$^-$ free-bound emission,
tiny protostars are formed in the clumps within their free-fall time
$\sim 10^3$ yr, which is an order of magnitude shorter than the
orbital decay time.  In this section, we simply estimate the accretion
rate on to the clumps (which we assume quickly evolve to protostars),
and then discuss the possibility of forming ZAMS stars from the
clumps.

The accretion rate on to a point-like clump in a Keplerian disc,
where we assume a rotationally supported disc \citep{2014MNRAS.439.1160R},
is estimated as
\begin{equation}
\mdot_{\rm c}=\frac{3}{2}\Sigma \Omega (f_{\rm H}R_{\rm H})^2,
\end{equation}
where $R_{\rm H}$ is the Hill radius defined by $R(M_{\rm
  c}/3M_\ast)^{1/3}$ and $f_{\rm H}\sim O(1)$
\citep[e.g.,][]{2004ApJ...608..108G}.  We have also assumed that the
clump accretes gas orbiting in the nearby disc, within an impact
parameter $f_{\rm H}R_{\rm H}$.
At the fragmentation radius, the accretion rate is given by
\begin{align}
\mdot_{\rm c}|_{R_{\rm f}}&=\frac{3}{2}\Sigma_{\rm f} \Omega_{\rm f} f_{\rm H}^2 R_{\rm f}^2\left(\frac{M_c}{3M_\ast}\right)^{2/3}\label{eq:clump_acc}\\\nonumber
&\simeq 1.1\times 10^{-2}~\msunyr~\left(\frac{f_{\rm H}}{1.5}\right)^2\left(\frac{M_{\rm c}}{30~\msun}\right)^{2/3},
\end{align}
which is similar to the critical rate ($\approx4\times
10^{-3}~\msunyr$) at which the evolution of a protostar changes
qualitatively \citep{2001ApJ...561L..55O, 2003ApJ...589..677O}.  Below
the critical rate, the protostar grows to a usual ZAMS star.  Above
the critical rate, the protostar evolves instead to a structure with a
bloated envelope resembling a giant star \citep{2012ApJ...756...93H, 2013ApJ...778..178H}.
The accretion rate given by Eq.~(\ref{eq:clump_acc}) 
is only a factor of $\approx 2$ above the critical value. We
therefore consider the case in which some of the clumps grow at a
sub-critical rate and form massive ZAMS stars.

The protostar embedded in the clump begins to undergo Kelvin-Helmholtz
(KH) contraction, loosing energy by radiative diffusion.  After the
contraction, the star reaches a ZAMS star when the central temperature
increases to $\sim 10^8$K and hydrogen burning begins
\citep{2001ApJ...561L..55O, 2003ApJ...589..677O}. 
Below the critical rate, the time-scale of the KH contraction is estimated as 
\begin{align}
t_{\rm KH}&\simeq \frac{M_{\rm c}}{\mdot_{\rm c}}
\ga 10^4~{\rm yr}\left(\frac{M_{\rm c}}{30~\msun }\right)
\left(\frac{\mdot_{\rm c}}{3\times 10^{-3}~\msunyr}\right)^{-1}.
\end{align}
Thus, it is longer than
both the migration time for $M_\ast \leq 10^4~\msun$ and the orbital
period at the fragmentation radius.
Therefore, most clumps formed
promptly in the early (disc-dominated) phase are expected to migrate
and merge with the central protostar before reaching the massive ZAMS
stars ($M_\ast \leq 10^4~\msun$). On the other hand, clumps formed in
the late (central protostar-dominated) phase ($M_\ast >10^4~\msun$)
could survive and evolve to ZAMS stars.

\subsection{Radiative feedback}

As we have seen in \S~\ref{sec:kh}, the decay time of the clump orbit
becomes longer than the KH time during the late stage for
$M_\ast>10^4~\msun$.  In this case, the clump can contract by loosing
energy and reach the ZAMS star before migrating towards the central
protostar.  This, however, requires a central SMS to be already present.
On the other hand, even during the early stage for $M_\ast <10^4~\msun$, 
some clumps could survive for their KH times, because
clumps interact with each other, as well as with the disc, within the
fragmentation radius.  In a clumpy disc, these gravitational
interactions will make the orbital decay time longer or shorter
stochastically, and can also cause clumps to be temporarily ejected
from the disc
\citep{2005ApJ...630..152E,2012MNRAS.424..399G,2012ApJ...746..110Z,2013ApJ...777L..14F}.
Some of these clumps can evolve to ZAMS stars and emit strong UV
radiation, which could prevent the accretion on to the protostar and
on to the disc as a while.

We briefly estimate the number of the clumps which exist in the disc at a moment.
The corresponding range where the fragmentation occurs is $\Delta R\simeq R_{\rm f}$.  
The Hill radius of the clump at $R_{\rm f}$ is written by
\begin{equation}
R_{\rm H}=1.5\times 10^{16}~{\rm cm}\left(\frac{M_{\rm c}}{30~\msun}\right)^{1/3}.
\label{eq:hill}
\end{equation}
Thus, we roughly estimate the maximum number of the clumps which survives as
\begin{equation}
N_{\rm c}\sim \frac{\Delta R}{R_{\rm H}}\simeq 20\left(\frac{M_\ast}{10^5~\msun}\right)^{1/3}
\left(\frac{M_{\rm c}}{30~\msun}\right)^{-1/3}.
\label{eq:nc}
\end{equation}
If some clumps grow at rates higher than $\sim 10^{-3}~\msunyr$ and
their masses increase within $t_{\rm KH}$, the maximum number
decreases.  
We here consider $N_{\rm c}\approx10-20$ as a conservative value.

\begin{figure}
\begin{center}
\includegraphics[height=36mm,width=83mm]{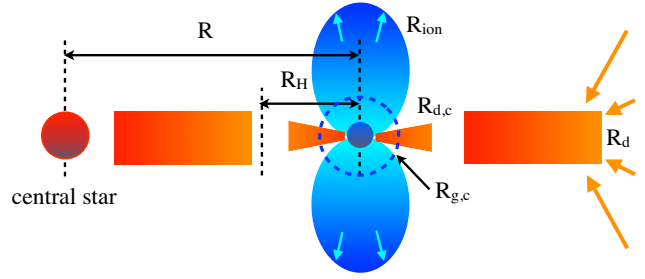}
\end{center}
\caption{Schematic figure of the accretion disc and a small
  circum clump disc, showing the distance of the clump from the
  central star $R$, the Hill radius of the clump $R_{\rm H}$ (Eq.~\ref{eq:hill}), 
  the size of the circum clump disc $R_{\rm d,c}$, the
  size of the H$_{\rm II}$ region around the clump $R_{\rm ion}$
  (Eq.~\ref{eq:ion}), the gravitational radius of the clump
  $R_{\rm g,c}$ (Eq.~\ref{eq:grav}), and the size of the whole
  nuclear disc $R_{\rm d}$ (Eq.~\ref{eq:rd}).  }
\label{fig:clump_disc}
\end{figure}

First, let us consider radiative feedback on the accretion on to the
ZAMS star (clump) itself.  Figure~\ref{fig:clump_disc} illustrates the
gas structure around the clumps and the characteristic radii.  The gas
accreting on to the ZAMS star makes a small circum-clump disc.  The
disc size is roughly estimated as $R_{\rm d,c}\simeq R_{\rm H}/3$
\citep{1998ApJ...508..707Q}, based on numerical simulations
\citep{2009MNRAS.397..657A, 2012ApJ...747...47T}.  The ZAMS star emits
UV radiation and an H$_{\rm II}$ region is formed above and below the
disc.  The size of the H$_{\rm II}$ region depends on the ionizing
luminosity of the ZAMS star and on the density profile of the
in-falling material.  For simplicity, we estimate the size $R_{\rm
  ion}$ in the polar direction using Eq.~(36) of
\cite{2008ApJ...681..771M} as
\begin{equation}
\frac{R_{\rm ion}}{R_{\rm d,c}}\simeq 0.16\left(\frac{M_{\rm c}}{30~\msun}\right)^{1.25}
\left(\frac{\mdot_{\rm c}}{3\times 10^{-3}~\msunyr}\right)^{-1},
\end{equation}
where we assume that the disc mass is equal to the clump mass and we
take the temperature of the H$_{\rm II}$ region to be $3\times 10^4$ K
\citep{2011Sci...334.1250H}.  The sound speed in the H$_{\rm II}$
region is $c_{\rm s,ion}\simeq 20$ km s$^{-1}$.  For a ZAMS star
located at $R_{\rm f}$, we obtain
\begin{equation}
R_{\rm ion}\simeq 8.0\times 10^{14}~{\rm cm} \left(\frac{M_{\rm c}}{30~\msun}\right)^{19/12}
\left(\frac{\mdot_{\rm c}}{3\times 10^{-3}~\msunyr}\right)^{-1}.
\label{eq:ion}
\end{equation}
Since the sound-crossing time within $R_{\rm ion}$ is much shorter
($\sim13$ yr), the ionization front expands rapidly.  When the front
reaches the gravitational radius
\begin{equation}
R_{\rm g,c}=\frac{GM_{\rm c}}{c_{\rm s,ion}^2}
\simeq 1.0\times 10^{15}~{\rm cm} \left(\frac{M_{\rm c}}{30~\msun}\right),
\label{eq:grav}
\end{equation}
the ionized gas breaks out through the neutral infalling gas.  The
ionizing photons subsequently heat the disc surface, and thus
photo-evaporation begins to suppress the accretion rate.  The
photo-evaporation rate can be expressed as
\begin{align}
\mdot_{\rm PE}\simeq 3.8\times 10^{-4}\left(\frac{\Phi_{\rm EUV}}{10^{50}~{\rm s}^{-1}}\right)^{1/2}
\left(\frac{R_{\rm d,c}}{10^{16}~{\rm cm}}\right)^{1/2}~\msunyr,
\end{align}
where $\Phi_{\rm EUV}$ is the ionizing photon number flux
\citep{2013ApJ...773..155T}.  Using the relation $\Phi_{\rm
  EUV}=3.7\times 10^{49}N_{\rm c}$ s$^{-1}(M_{\rm c}/60~\msun)^{3/2}$
in the mass range $60\la M_{\rm c} \la 300~\msun$, we obtain
\begin{align}
\mdot_{\rm PE}\simeq 7.4\times 10^{-4}\left(\frac{M_\ast}{10^5~\msun}\right)^{1/6}
\left(\frac{M_{\rm c}}{60~\msun}\right)^{3/4}
~\msunyr,
\end{align}
which is smaller than $\mdot_{\rm c}\sim {\rm a~few}\times
10^{-3}~\msunyr$ by approximately an order of magnitude.  We therefore
conclude that accretion on to the clump could not be suppressed by its
own photoionization heating.  After the H$_{\rm II}$ region breaks out
of the disc, however, the accretion proceeds only from the shadow of
the disc and thus the accretion rate is reduced.  Moreover,
\cite{2011Sci...334.1250H} suggest that the gas behind the disc is
shocked and accelerated outward by the pressure gradient after the
expansion of the H$_{\rm II}$ region.  This process halts the
accretion at $M_{\rm c} \approx 40~\msun$.  
\citet{2013ApJ...773..155T}
have shown that photo-evaporation starts to suppress the accretion 
when $\mdot_{\rm PE}/\mdot_{\rm c}\ga 0.2$.
Note that accretion on to a clump inside $R_{\rm f}$ cannot be
suppressed, even if the H$_{\rm II}$ region expands by a large factor.
In particular, a clump growing at a super critical rate, $\mdot_{\rm
  c}\ga 10^{-2}~\msunyr$, is not affected by the radiative feedback
because $\mdot_{\rm PE}/\mdot_{\rm c} \la 0.08$.

Next, we discuss the possibility that the ionizing photons from the
collection of all ZAMS stars together suppress the accretion (with
$\mdot_{\rm tot}\simeq 0.1~\msunyr$) from the parent cloud on to the
disc as a whole.  To make our discussion conservative, we assume that
the clumps which emit strong UV radiation grow at the rate just below
the critical accretion rate.  For $M_\ast \sim 10^5~\msun$, the
orbital decay time is $\sim 3\times 10^4$ yr and so each clump then
grows to $\sim 100~\msun$.  The corresponding total ionizing photon
rate is $\sim 1.1\times 10^{51}$~s$^{-1}$.  From numerical simulations
of the gravitational collapse of an atomic-cooling cloud
\citep{2014arXiv1404.4630I}, the rotational velocity is proportional
to the Keplerian velocity, $v_{\rm rot}=f_{\rm Kep}v_{\rm Kep}$, with
$f_{\rm Kep}\simeq 0.5$.  From this relation, the size of the whole
disc around the central protostar is given by
\begin{align}
R_{\rm d}&=f_{\rm Kep}^2 R_{\rm env},\label{eq:rd}\\\nonumber
&\simeq 2.5~{\rm pc}\left(\frac{f_{\rm Kep}}{0.5}\right)^{2}
\left(\frac{R_{\rm env}}{10~{\rm pc}}\right),
\end{align}
where $R_{\rm env}$ is the size of the quasi-spherical parent cloud
fuelling the inner disc. We approximate $R_{\rm env}$ using a critical
Bonnor-Ebert sphere with a temperature of $8,000$ K and a central
density of $10^4~\cc$, 
which corresponds to that of the gravitationally unstable core 
in the atomic cooling haloes without H$_2$ molecules \citep[e.g.,][]{2008ApJ...682..745W}.
Therefore, we obtain $\mdot_{\rm PE}=3.5\times 10^{-2}~\msunyr~(\Phi_{\rm EUV}/1.1\times 10^{51}~{\rm s}^{-1})^{1/2}
(R_{\rm d}/2.5~{\rm pc})^{1/2}$ and thus $\mdot_{\rm PE}/\mdot_{\rm tot}\simeq 0.35$,
which is comparable or only slightly above the critical value.
Therefore, even with conservative assumptions, the strong UV radiation
cannot deplete the gas supply on to the disc from the parent cloud, and
the central protostar can grow to a SMS star with $M_\ast \simeq 10^5~\msun$.


\section{fragmentation by metals: dust cooling}
\label{sec:metal}

The site envisioned for forming a SMS is atomic cooling gas in a halo
with virial temperature $>10^4$ K, in which H$_2$ cooling is
prohibited prior to and throughout the protostellar collapse.  Recent
numerical simulations suggest that the majority of the atomic cooling
haloes are polluted by heavy elements due to Pop III supernovae from
prior star formation \citep{2010ApJ...716..510G,2012ApJ...745...50W}.
The resulting metallicity is $Z\la 10^{-4}~\zsun$ ($\la
10^{-3}~\zsun$) if the Pop III supernovae are core-collapse type
(pair-instability type).  If the gas is slightly polluted with $Z\ga
5\times 10^{-6}~\zsun$, the temperature decreases below $\sim 500$ K
by dust cooling \citep{2008ApJ...686..801O}, which could promote
efficient fragmentation.  The details of this fragmentation, and
whether dust cooling ultimately prevents SMS formation, are not yet
understood.

The temperature of the gas with dust grains begins to decrease through
heat exchange with cool dust grains ($T_{\rm gr}\ll T$) when collisions
between gas particles and dust are sufficiently frequent.  Collisional
cooling of the gas is efficient until $T\sim 500$ K, where the gas and
dust is thermally coupled ($T\simeq T_{\rm gr}$).  At the cooling
phase, the compressional heating and collisional cooling are balanced
as
\begin{equation}
\frac{\rho c_{\rm s}^2}{t_{\rm ff}}\simeq H_{\rm gr},
\label{eq:d1}
\end{equation}
where $t_{\rm ff}=\sqrt{3\pi /(32G\rho)}$ is the free-fall time and
$H_{\rm gr}$ is the energy exchange rate (in erg s$^{-1}$ cm$^{-3}$)
between the gas and dust, defined by
\begin{equation}
H_{\rm gr}\simeq 2k_{\rm B}Tn_{\rm H}n_{\rm gr}\sigma_{\rm gr}c_{\rm s},
\label{eq:d2}
\end{equation}
for $T\gg T_{\rm gr}$ \citep{2006MNRAS.369.1437S,2012MNRAS.419.1566S}.
We adopt the number density and cross-section of the dust particles
from \cite{2003A&A...410..611S},
\begin{equation}
\frac{n_{\rm gr}\sigma_{\rm gr}}{\rho}=4.7\times 10^{-2}\left(\frac{Z}{10^{-4}~\zsun}\right)~{\rm cm^2~g^{-1}},
\label{eq:d3}
\end{equation}
where the depletion factor of metals to dusts is assumed as high as the present-day Galactic value 
$f_{\rm dep}\simeq 0.5$.
From Eqs.~(\ref{eq:d1}), (\ref{eq:d2}), and (\ref{eq:d3}), we obtain 
\begin{equation}
n\simeq 6.8\times 10^7~\cc \left(\frac{T}{6000~{\rm K}}\right)^{-1}\left(\frac{Z}{10^{-4}~\zsun}\right)^{-2},
\label{eq:d4}
\end{equation}
above which the temperature begins to decrease compared to the
zero-metallicity case.  The density given by Eq.~(\ref{eq:d4}) is
comparable to that at the fragmenting radius.  This implies that the
discussion in \S~\ref{sec:fate} remains valid in metal-polluted gas,
as long as the metallicity remains below $Z\la 10^{-4}~\zsun$.  We
conclude that disc fragmentation cannot prevent SMS formation, as long
as the metallicity is below this value. 
On the other hand, at
higher metallicity ($Z> 3\times10^{-4}~\zsun$),
the gas temperature rapidly decreases by metal-line cooling at
densities below $\sim 10^4~\cc$ \citep{2008ApJ...686..801O,2012MNRAS.422.2539I}.  
Numerical simulations of metal-enriched collapse into atomic cooling haloes
also capture the character of gas fragmentation for $Z\ga 10^{-3}~\zsun$
\citep{2014MNRAS.438.1669S,2014MNRAS.440L..76S}.
The outcome would be
a compact star-cluster which consists of many low-mass stars with
$\sim 1~\msun$ \citep{2014MNRAS.439.1884T}.


\section{Discussion and conclusions}
\label{sec:conc}

In this paper, we discussed the properties of a disc around the embryo
of a SMS ($\ga 10^5~\msun$), expected to be present
in a primordial gas without H$_2$ molecules in massive haloes with
virial temperature $\ga 10^4$ K.  A high accretion rate $\ga
0.1~\msunyr$, sustained for $\ga 10^5$yr, is required to form a SMS.
The inner region of such a disc is gravitationally unstable, and
fragments into $O(10)$ clumps with characteristic mass of $\sim
30~\msun$.  We discuss the possibility that this fragmentation
prevents SMS formation.  We argue that most of the clumps formed in
the disc rapidly migrate towards the central protostar and merge with
it.  The orbital decay time is shorter than or comparable to the
orbital period of the clumps ($\la 10^4$ yr).  Some of the clumps can
grow via accretion and evolve to ZAMS stars within
their KH time, either because they survive longer
due to the stochasticity of the migration process, or because they
form later, further out in the disc.

Our toy model, on which these conclusions are based,
  can be tested against simulations of Pop III star formation in
  lower mass minihaloes.  In this case, the dominant cooling process is
  H$_2$ line emission ($T\sim 10^3$ K) and the resulting accretion
  rate is $\sim 10^{-3}-10^{-2}~\msunyr$, which is smaller than that
  in the SMS formation case.  According to high-resolution numerical
  simulations by \cite{2012MNRAS.424..399G}, clump formation occurs at
  $\sim 10$ AU ($\Sigma \sim 5\times 10^3$ g cm$^{-2}$, $\Omega \sim
  0.1$ yr$^{-1}$, $c_{\rm s}\sim 2.5$ km s$^{-1}$ and $M_{\rm d}\sim
  1~\msun$ in their Fig.~2) in the gravitationally unstable disc at
  the early stage $\la 10$ yr.  From Eq.~(\ref{eq:r_f_early}), we can
  estimate the fragmentation radius as $\sim 20$ AU, which is
  consistent with the fragmentation radii in the simulations.  At this
  stage, the average accretion rate is high, $\mdot_{\rm tot}\simeq
  10^{-2}~\msunyr$ in their Fig.~8.  From Eq.~(\ref{eq:continue}),
  (\ref{eq:scaleheight}), (\ref{eq:viscous}), and $Q\simeq 1$,
\begin{equation}
\mdot_{\rm tot}\simeq 3\alpha \frac{c_{\rm s}^3}{G}.
\end{equation}
Using these equations, the sound speed at the fragmentation radius
($\alpha _{\rm f}=1$) is estimated as $c_{\rm s}=2.5$ km s$^{-1}$, which
agrees well with the numerical result.  The viscous time at $\sim 10$
AU is roughly estimated as $t_{\rm vis}\sim M_{\rm d}/\mdot_{\rm
  tot}\sim 10^2$ yr.  Since some clumps migrate inward within $10$ yr
($< t_{\rm vis}$) in simulation, other effects (Type I migration and
interaction between clumps) could accelerate the clumps migration.
Therefore, our results from the toy model are expected to be
conservative, and appear to capture the essential features of disc
fragmentation and clump migration.

The gas in most of the massive haloes is likely to be polluted by heavy
elements due to prior star formation and Pop III supernovae.  With the
metallicity of $5\times 10^{-6} \la Z \la 10^{-4}~\zsun$, the dust
cooling decreases the temperature rapidly and could promote the
fragmentation at $n> 10^{8}~\cc$ \citep{2008ApJ...686..801O}.  As a
result, the existence of dusts has been considered as one of the most
severe obstacles to the SMS formation.  However, the fragmentation
induced by the dust cooling is expected to occur only within the
fragmenting radius we estimated. 
This means that the clumps formed by 
the dust cooling can migrate towards the center on a time-scale comparable to the 
orbital period, in the same way as the primordial case. We conclude
that the SMS formation is not prevented, unless the metal-line cooling
dominates for $Z\ga 3\times 10^{-4}~\zsun$.  Our results therefore
remove a significant obstacle for SMS formation.

Our estimates for the orbital decay time of the clumps are based on
the well-known formulae of Type~I and II planetary migration.
Although some numerical simulations find that these formulae remain
approximately valid in gravitationally unstable and clumpy discs,
there are many uncertainties regarding the migration rate in this
case.  For example, we neglect the orbital eccentricities of the
clumps.  The clumps formed by the disc fragmentation are expected to
have eccentric orbits and thus the angular momentum transfer between
the disc and clumps could change from the case of the circular orbits.
Moreover, in such a situation, the interaction with eccentric clumps
retard the orbital decay of the other clumps
\citep{2013ApJ...777L..14F}.  To estimate the migration time within
the clumpy disc around the SMS star, more sophisticated numerical
simulations are necessary.

It is worth further emphasizing the limitations and uncertainties of
our disc model.  In Figure~\ref{fig:profile}, we present the disc
profile based on cooling by free-bound emission of H$^-$ and thermal
equilibrium.  Hydrogen atomic cooling (Ly$\alpha$ and two-photon
emission) dominates at $T\ga 8000$ K ($P>10^4$ yr).  To quantify the
effect, we include Ly$\alpha$ cooling ($\Lambda =\lambda_{\rm
  Ly\alpha}n^2x_{\rm e}$; \citealt{2007ApJ...666....1G}) in the
limiting optically thin case (dashed lines in Fig.~\ref{fig:profile}).
Since the gas is in fact optically thick to the Ly$\alpha$ photons at
$n>10^5~\cc$, and the cooling efficiency of the two-photon channel is
smaller than the Ly$\alpha$ emission, the dashed lines overestimate
the additional atomic cooling rate.  Nevertheless, the profiles
deviate from our fiducial disc model only modestly, beginning at
$P>10^4$ yr and thus the critical period changes at most within the
shaded region in Figure~\ref{fig:profile} (c).

The crucial ingredient of this paper is the critical value of the
effective viscous parameter $\alpha_{\rm f}$ for fragmentation.  In
our model, the fragmentation radius depends sensitively on the value
of $\alpha_{\rm f}$.  For example, if we chose $\alpha_{\rm f}=0.5$,
we obtain $\Omega_{\rm f}=2\times 10^{-12}$ s$^{-1}$ ($P=10^5$ yr),
$\Sigma_{\rm f}=6$ g cm$^{-2}$, 
$M_{\rm c}\simeq 400~\msun$ and $R_{\rm f}\simeq 0.2$ pc.  In this
case, the orbital decay time becomes longer than the KH time of the
clumps by a factor of 4, and most of the clumps may evolve to
massive stars and emit strong UV radiation. 
Nevertheless, the number
of clumps decreases by a factor of 2 from the case of $\alpha_{\rm f}=1.0$
(see Eq. \ref{eq:nc}).
Therefore, we expect that our main conclusion also does not change. 
However, if $\alpha_{\rm f}\la 0.1$, the fragmenting radius moves outside the size of the
whole disc given by Eq.~(\ref{eq:rd}), which means that the disc
cannot exist, i.e., our model is no longer self-consistent.  To
explore the precise value of $\alpha_{\rm f}$ is left for future
investigations.  In particular, our results motivate high-resolution
numerical simulations similar to \cite{2014MNRAS.439.1160R}.
Moreover, more realistic treatments of radiative cooling and chemical
reactions at densities higher than $\sim 10^8~\cc$
\citep{2014arXiv1404.4630I} should be included since the fragmentation
efficiency strongly depends on the equation of state of the gas.

Despite these caveats, we expect that SMS formation in metal-poor gas
in atomic-cooling haloes is difficult to avoid, as it requires that
the high mass inflow rate, $\ga 0.1~\msunyr$, is democratically
distributed among $O(100)$ fragments (so that {\em none} of them
accretes at a super critical rate for SMS formation) and that most of
these fragments survive rapid migration and avoid coalescence with the
growing central protostar for several orbital times.  The toy models
presented in this paper disfavor this scenario.

\section*{Acknowledgements} 
We thank Greg Bryan, Kazuyuki Omukai, Takashi Hosokawa, Kei Tanaka and
Eli Visbal for fruitful discussions.  This work is supported by the
Grants-in-Aid by the Ministry of Education, Culture, and Science of
Japan (KI), and by NASA grant NNX11AE05G (ZH).

\appendix
\section{ionization degree}
\label{sec:appA}
In this appendix, we briefly describe how to obtain the expression of
Eq.~(\ref{eq:x_e}).  The reaction rate coefficient of radiative
recombination $\alpha_{\rm rec}$ is a function of temperature \citep{1992ApJ...387...95F}.
On the other hand, that of the collisional ionization $\alpha_{\rm ci}$ has a complex form 
\citep{2001ApJ...546..635O}
\begin{equation}
\alpha_{\rm ci}=\alpha_{\rm inv}
\left(\frac{z_{\rm H_2^+}z_{\rm e}}{z_{\rm H}^2}\right)~
3.6034\times 10^{-5}~\exp(-\epsilon /T),
\end{equation}
and
\begin{equation}
\alpha_{\rm inv}=1.32\times 10^{-6}T^{-0.76}f_{\rm 1st,ex},
\end{equation}
where $z_{\rm H(H_2^+,~e)}$ are the partition functions, $\epsilon =1.2713\times 10^5$ K, and
$f_{\rm 1st,ex}$ is defined by
\begin{equation}
f_{\rm 1st,ex}=\frac{n}{n+n_{\rm H,cr}},
\end{equation}
and
\begin{equation}
\frac{n}{n_{\rm H,cr}}=\frac{C_{21}}{A_{21}\beta_{21}+A_{\rm 2ph}}
=\frac{\gamma_{\rm H,21}x_{\rm H}+
\gamma_{\rm e, 21}x_{\rm e}}{A_{21}\beta_{21}+A_{\rm 2ph}}n,
\end{equation}
where $C_{21}$ is the collisional transition rate,
$A_{21}$ ($A_{\rm 2ph}$) the radiative transit rate from the first excited state to the ground state 
(two-photon emission, respectively),  and $\beta_{21}$ the escape probability of photons with $10.2$ eV
(see \citealt{2001ApJ...546..635O} in details).
For a wide range of densities ($10^6<n<10^{14}$ cm$^{-3}$), the
following approximation is valid:
\begin{align}
f_{\rm 1st,ex}&\approx \frac{n}{n_{\rm H,cr}} \approx
\frac{\gamma_{\rm H,21}x_{\rm H}}{A_{\rm 2ph}}n\\\nonumber
&\approx 4.3\times 10^{-16}~n~T^{0.57}~
(500<T<10^4~{\rm K}).
\end{align}


\begin{thebibliography}{99}
{\small


\bibitem[Agarwal et al.(2012)]{2012MNRAS.425.2854A} Agarwal, B., Khochfar, 
S., Johnson, J.~L., et al.\ 2012, MNRAS, 425, 2854 


\bibitem[Alvarez et al.(2009)]{2009ApJ...701L.133A} Alvarez, M.~A., Wise, 
J.~H., \& Abel, T.\ 2009, ApJL, 701, L133 


\bibitem[Ayliffe 
\& Bate(2009)]{2009MNRAS.397..657A} Ayliffe, B.~A., \& Bate, M.~R.\ 2009, MNRAS, 397, 657 


\bibitem[Baruteau et al.(2011)]{2011MNRAS.416.1971B} Baruteau, C., Meru, 
F., \& Paardekooper, S.-J.\ 2011, MNRAS, 416, 1971 


\bibitem[Begelman et al.(2006)]{2006MNRAS.370..289B} Begelman, M.~C., 
Volonteri, M., \& Rees, M.~J.\ 2006, MNRAS, 370, 289 


\bibitem[Bromm 
\& Loeb(2003)]{2003ApJ...596...34B} Bromm, V., \& Loeb, A.\ 2003, ApJ, 596, 34 


\bibitem[Ceverino et al.(2010)]{2010MNRAS.404.2151C} Ceverino, D., Dekel, 
A., \& Bournaud, F.\ 2010, MNRAS, 404, 2151 


\bibitem[Clark et al.(2011)]{2011Sci...331.1040C} Clark, P.~C., Glover, 
S.~C.~O., Smith, R.~J., et al.\ 2011, Science, 331, 1040 


\bibitem[Crida et al.(2006)]{2006Icar..181..587C} Crida, A., Morbidelli, 
A., \& Masset, F.\ 2006, Icarus, 181, 587 


\bibitem[Dijkstra et al.(2008)]{2008MNRAS.391.1961D} Dijkstra, M., Haiman, 
Z., Mesinger, A., \& Wyithe, J.~S.~B.\ 2008, MNRAS, 391, 1961 


\bibitem[Dijkstra et al.(2014)]{2014MNRAS.442.2036D} Dijkstra, M., Ferrara, 
A., \& Mesinger, A.\ 2014, MNRAS, 442, 2036 


\bibitem[Di Matteo et al.(2012)]{2012ApJ...745L..29D} Di Matteo, T., 
Khandai, N., DeGraf, C., et al.\ 2012, ApJL, 745, L29 


\bibitem[Duffell \& MacFadyen(2013)]{duffellmacfadyen2013} Duffell, P. C., \& MacFadyen, A. I. 2013, ApJ, 769, 41


\bibitem[Duffell et al.(2014)]{duffell+2014} Duffell, P.~C., Haiman, 
Z., MacFadyen, A.~I., D'Orazio, D.~J., 
\& Farris, B.~D.\ 2014, ApJL, 792, L10 


\bibitem[Escala et al.(2005)]{2005ApJ...630..152E} Escala, A., Larson, 
R.~B., Coppi, P.~S., \& Mardones, D.\ 2005, ApJ, 630, 152 

\bibitem[Escala(2007)]{2007ApJ...671.1264E} Escala, A.\ 2007, ApJ, 671, 
1264 

\bibitem[Fan(2006)]{2006NewAR..50..665F} Fan, X.\ 2006, Nature, 50, 665 


\bibitem[Fiacconi et al.(2013)]{2013ApJ...777L..14F} Fiacconi, D., Mayer, 
L., Ro{\v s}kar, R., \& Colpi, M.\ 2013, ApJL, 777, L14 


\bibitem[Ferland et al.(1992)]{1992ApJ...387...95F} Ferland, G.~J., 
Peterson, B.~M., Horne, K., Welsh, W.~F., 
\& Nahar, S.~N.\ 1992, ApJ, 387, 95 


\bibitem[Fernandez et al.(2014)]{2014MNRAS.439.3798F} Fernandez, R., Bryan, 
G.~L., Haiman, Z., \& Li, M.\ 2014, MNRAS, 439, 3798 


\bibitem[Gammie(2001)]{2001ApJ...553..174G} Gammie, C.~F.\ 2001, ApJ, 553, 
174 


\bibitem[Goldreich 
\& Tremaine(1980)]{1980ApJ...241..425G} Goldreich, P., \& Tremaine, S.\ 1980, ApJ, 241, 425 


\bibitem[Goodman 
\& Tan(2004)]{2004ApJ...608..108G} Goodman, J., \& Tan, J.~C.\ 2004, ApJ, 608, 108 


\bibitem[Greif et al.(2010)]{2010ApJ...716..510G} Greif, T.~H., Glover, 
S.~C.~O., Bromm, V., \& Klessen, R.~S.\ 2010, ApJ, 716, 510 

\bibitem[Greif et al.(2011)]{greif+11} Greif, T.~H., Springel, V., White, S. D. M., Glover, S. C. O., Clark, P. C., Smith, R. J., Klessen, R. S., \& Bromm, V.\ 2011, ApJ, 737, 75

\bibitem[Greif et al.(2012)]{2012MNRAS.424..399G} Greif, T.~H., Bromm, V., 
Clark, P.~C., et al.\ 2012, MNRAS, 424, 399 


\bibitem[Glover 
\& Jappsen(2007)]{2007ApJ...666....1G} Glover, S.~C.~O., \& Jappsen, A.-K.\ 2007, ApJ, 666, 1 


\bibitem[Haiman 
\& Loeb(2001)]{2001ApJ...552..459H} Haiman, Z., \& Loeb, A.\ 2001, ApJ, 552, 459 


\bibitem[Hosokawa et al.(2011)]{2011Sci...334.1250H} Hosokawa, T., Omukai, 
K., Yoshida, N., \& Yorke, H.~W.\ 2011, Science, 334, 1250 


\bibitem[Hosokawa et al.(2012)]{2012ApJ...756...93H} Hosokawa, T., Omukai, 
K., \& Yorke, H.~W.\ 2012, ApJ, 756, 93 


\bibitem[Hosokawa et al.(2013)]{2013ApJ...778..178H} Hosokawa, T., Yorke, 
H.~W., Inayoshi, K., Omukai, K., \& Yoshida, N.\ 2013, ApJ, 778, 178 


\bibitem[Inayoshi 
\& Omukai(2011)]{2011MNRAS.416.2748I} Inayoshi, K., \& Omukai, K.\ 2011, MNRAS, 416, 2748 


\bibitem[Inayoshi 
\& Omukai(2012)]{2012MNRAS.422.2539I} Inayoshi, K., \& Omukai, K.\ 2012, MNRAS, 422, 2539 


\bibitem[Inayoshi et al.(2014)]{2014arXiv1404.4630I} Inayoshi, K., Omukai, 
K., \& Tasker, E.~J.\ 2014, MNRAS, submitted (arXiv:1404.4630)


\bibitem[Johnson 
\& Bromm(2007)]{2007MNRAS.374.1557J} Johnson, J.~L., \& Bromm, V.\ 2007, MNRAS, 374, 1557 


\bibitem[Johnson et al.(2013)]{2013MNRAS.428.1857J} Johnson, J.~L., Dalla 
Vecchia, C., \& Khochfar, S.\ 2013, MNRAS, 428, 1857 


\bibitem[Krumholz et al.(2007)]{2007ApJ...656..959K} Krumholz, M.~R., 
Klein, R.~I., \& McKee, C.~F.\ 2007, ApJ, 656, 959 


\bibitem[Latif et al.(2013)]{2013MNRAS.433.1607L} Latif, M.~A., Schleicher, 
D.~R.~G., Schmidt, W., \& Niemeyer, J.\ 2013, MNRAS, 433, 1607 


\bibitem[Levin(2007)]{2007MNRAS.374..515L} Levin, Y.\ 2007, MNRAS, 374, 
515 


\bibitem[Li et al.(2007)]{2007ApJ...665..187L} Li, Y., Hernquist, L., 
Robertson, B., et al.\ 2007, ApJ, 665, 187 


\bibitem[Lin 
\& Papaloizou(1979)]{1979MNRAS.186..799L} Lin, D.~N.~C., \& Papaloizou, J.\ 1979, MNRAS, 186, 799 


\bibitem[Lin 
\& Papaloizou(1986)]{1986ApJ...309..846L} Lin, D.~N.~C., \& Papaloizou, J.\ 1986, ApJ, 309, 846 


\bibitem[Lodato 
\& Natarajan(2006)]{2006MNRAS.371.1813L} Lodato, G., \& Natarajan, P.\ 2006, MNRAS, 371, 1813 


\bibitem[Loeb 
\& Rasio(1994)]{1994ApJ...432...52L} Loeb, A., \& Rasio, F.~A.\ 1994, ApJ, 432, 52 


\bibitem[McKee 
\& Tan(2008)]{2008ApJ...681..771M} McKee, C.~F., \& Tan, J.~C.\ 2008, ApJ, 681, 771 


\bibitem[Meru 
\& Bate(2011a)]{2011MNRAS.410..559M} Meru, F., \& Bate, M.~R.\ 2011a, MNRAS, 410, 559 


\bibitem[Meru 
\& Bate(2011b)]{2011MNRAS.411L...1M} Meru, F., \& Bate, M.~R.\ 2011b, MNRAS, 411, L1 


\bibitem[Michael et al.(2011)]{2011ApJ...737L..42M} Michael, S., Durisen, 
R.~H., \& Boley, A.~C.\ 2011, ApJL, 737, L42 


\bibitem[Milosavljevi{\'c} et al.(2009)]{2009ApJ...696L.146M} 
Milosavljevi{\'c}, M., Couch, S.~M., \& Bromm, V.\ 2009, ApJL, 696, L146 


\bibitem[Mortlock et al.(2011)]{2011Natur.474..616M} Mortlock, D.~J., 
Warren, S.~J., Venemans, B.~P., et al.\ 2011, Nature, 474, 616 


\bibitem[Omukai(2001)]{2001ApJ...546..635O} Omukai, K.\ 2001, ApJ, 546, 
635 


\bibitem[Omukai 
\& Palla(2001)]{2001ApJ...561L..55O} Omukai, K., \& Palla, F.\ 2001, ApJL, 561, L55 


\bibitem[Omukai 
\& Palla(2003)]{2003ApJ...589..677O} Omukai, K., \& Palla, F.\ 2003, ApJ, 589, 677 


\bibitem[Omukai et al.(2008)]{2008ApJ...686..801O} Omukai, K., Schneider, 
R., \& Haiman, Z.\ 2008, ApJ, 686, 801 


\bibitem[Park 
\& Ricotti(2012)]{2012ApJ...747....9P} Park, K., \& Ricotti, M.\ 2012, ApJ, 747, 9 


\bibitem[Quillen 
\& Trilling(1998)]{1998ApJ...508..707Q} Quillen, A.~C., \& Trilling, D.~E.\ 1998, ApJ, 508, 707 


\bibitem[Rafikov(2002)]{2002ApJ...572..566R} Rafikov, R.~R.\ 2002, ApJ, 
572, 566 


\bibitem[Rafikov(2005)]{2005ApJ...621L..69R} Rafikov, R.~R.\ 2005, ApJL, 
621, L69 


\bibitem[Regan 
\& Haehnelt(2009)]{2009MNRAS.396..343R} Regan, J.~A., \& Haehnelt, M.~G.\ 2009, MNRAS, 396, 343 


\bibitem[Regan et al.(2014)]{2014MNRAS.439.1160R} Regan, J.~A., Johansson, 
P.~H., \& Haehnelt, M.~G.\ 2014, MNRAS, 439, 1160 


\bibitem[Reisswig et al.(2013)]{2013PhRvL.111o1101R} Reisswig, C., Ott, 
C.~D., Abdikamalov, E., et al.\ 2013, Physical Review Letters, 111, 151101 


\bibitem[Rice et al.(2005)]{2005MNRAS.364L..56R} Rice, W.~K.~M., Lodato, 
G., \& Armitage, P.~J.\ 2005, MNRAS, 364, L56 


\bibitem[Safranek-Shrader et al.(2014a)]{2014MNRAS.438.1669S} 
Safranek-Shrader, C., Milosavljevi{\'c}, M., 
\& Bromm, V.\ 2014, MNRAS, 438, 1669 


\bibitem[Safranek-Shrader et al.(2014b)]{2014MNRAS.440L..76S} 
Safranek-Shrader, C., Milosavljevi{\'c}, M., 
\& Bromm, V.\ 2014, MNRAS, 440, L76 


\bibitem[Schneider et al.(2006)]{2006MNRAS.369.1437S} Schneider, R., 
Omukai, K., Inoue, A.~K., \& Ferrara, A.\ 2006, MNRAS, 369, 1437 


\bibitem[Schneider et al.(2012)]{2012MNRAS.419.1566S} Schneider, R., 
Omukai, K., Bianchi, S., \& Valiante, R.\ 2012, MNRAS, 419, 1566 


\bibitem[Semenov et 
al.(2003)]{2003A&A...410..611S} Semenov, D., Henning, T., Helling, C., Ilgner, M., \& Sedlmayr, E.\ 2003, A\&A, 410, 611 


\bibitem[Shakura 
\& Sunyaev(1973)]{1973A&A....24..337S} Shakura, N.~I., \& Sunyaev, R.~A.\ 1973, A\&A, 24, 337 


\bibitem[Shang et al.(2010)]{2010MNRAS.402.1249S} Shang, C., Bryan, G.~L., 
\& Haiman, Z.\ 2010, MNRAS, 402, 1249 


\bibitem[Shapiro 
\& Teukolsky(1983)]{1983bhwd.book.....S} Shapiro, S.~L., \& Teukolsky, S.~A.\ 1983, 
Black Holes, White Dwarfs, and Neutron Stars: The Physics
of Compact Objects. Wiley-Interscience, New York

\bibitem[Shibata 
\& Shapiro(2002)]{2002ApJ...572L..39S} Shibata, M., \& Shapiro, S.~L.\ 2002, ApJL, 572, L39 


\bibitem[Shu(1977)]{1977ApJ...214..488S} Shu, F.~H.\ 1977, ApJ, 214, 488 


\bibitem[Stahler et al.(1986)]{1986ApJ...302..590S} Stahler, S.~W., Palla, 
F., \& Salpeter, E.~E.\ 1986, ApJ, 302, 590 



\bibitem[Stacy et al.(2010)]{2010MNRAS.403...45S} Stacy, A., Greif, T.~H., 
\& Bromm, V.\ 2010, MNRAS, 403, 45 



\bibitem[Syer 
\& Clarke(1995)]{1995MNRAS.277..758S} Syer, D., \& Clarke, C.~J.\ 1995, MNRAS, 277, 758 


\bibitem[Tanaka et al.(2002)]{2002ApJ...565.1257T} Tanaka, H., Takeuchi, 
T., \& Ward, W.~R.\ 2002, ApJ, 565, 1257 


\bibitem[Tanaka \& Haiman(2009)]{th2009} Tanaka, T., \& Haiman, Z.\ 2009, ApJ, 696, 1798

\bibitem[Tanaka et al.(2012)]{2012MNRAS.425.2974T} Tanaka, T., Perna, R., 
\& Haiman, Z.\ 2012, MNRAS, 425, 2974 


\bibitem[Tanaka et al.(2013)]{2013ApJ...773..155T} Tanaka, K.~E.~I., 
Nakamoto, T., \& Omukai, K.\ 2013, ApJ, 773, 155 


\bibitem[Tanaka 
\& Omukai(2014)]{2014MNRAS.439.1884T} Tanaka, K.~E.~I., \& Omukai, K.\ 2014, MNRAS, 439, 1884 


\bibitem[Tanigawa et al.(2012)]{2012ApJ...747...47T} Tanigawa, T., Ohtsuki, 
K., \& Machida, M.~N.\ 2012, ApJ, 747, 47 


\bibitem[Toomre(1964)]{1964ApJ...139.1217T} Toomre, A.\ 1964, ApJ, 139, 
1217 


\bibitem[Volonteri et al.(2003)]{2003ApJ...582..559V} Volonteri, M., 
Haardt, F., \& Madau, P.\ 2003, ApJ, 582, 559 


\bibitem[Vorobyov 
\& Basu(2005)]{2005ApJ...633L.137V} Vorobyov, E.~I., \& Basu, S.\ 2005, ApJL, 633, L137 


\bibitem[Vorobyov 
\& Basu(2006)]{2006ApJ...650..956V} Vorobyov, E.~I., \& Basu, S.\ 2006, ApJ, 650, 956 


\bibitem[Ward(1986)]{1986Icar...67..164W} Ward, W.~R.\ 1986, Icarus, 67, 
164 


\bibitem[Ward(1997)]{1997Icar..126..261W} Ward, W.~R.\ 1997, Icarus, 126, 
261 


\bibitem[Willott et al.(2007)]{2007AJ....134.2435W} Willott, C.~J., 
Delorme, P., Omont, A., et al.\ 2007, AJ, 134, 2435 


\bibitem[Wise et al.(2008)]{2008ApJ...682..745W} Wise, J.~H., Turk, M.~J., 
\& Abel, T.\ 2008, ApJ, 682, 745 


\bibitem[Wise et al.(2012)]{2012ApJ...745...50W} Wise, J.~H., Turk, M.~J., 
Norman, M.~L., \& Abel, T.\ 2012, ApJ, 745, 50 


\bibitem[Wolcott-Green et al.(2011)]{2011MNRAS.418..838W} Wolcott-Green, 
J., Haiman, Z., \& Bryan, G.~L.\ 2011, MNRAS, 418, 838 


\bibitem[Zeldovich 
\& Novikov(1971)]{1971reas.book.....Z} Zeldovich, Y.~B., \& Novikov, I.~D.\ 1971, 
Relativistic Astrophysics, Vol. 1, Stars and
Relativity. University of Chicago Press, Chicago 


\bibitem[Zhu et al.(2012)]{2012ApJ...746..110Z} Zhu, Z., Hartmann, L., 
Nelson, R.~P., \& Gammie, C.~F.\ 2012, ApJ, 746, 110 

}
\end{thebibliography}
\end{document}